\documentclass[twocolumn]{aastex62}
\usepackage{amsmath,amssymb,epsfig,bm,mathrsfs}
\newcommand{\bea}{\begin{eqnarray}}
\newcommand{\eea}{\end{eqnarray}}

\newcommand{\vecxi}{{\bm \xi}}
\newcommand{\vecOmega}{{\bm \Omega}}
\newcommand{\vecomega}{{\bm \omega}}
\newcommand{\vecnabla}{{\bm \nabla}}
\newcommand{\vecsigma}{{\bm \sigma}}
\newcommand{\vecx}{{\bm x}}
\definecolor{red}{rgb}{0.8,0,0}
\definecolor{violet}{rgb}{0.4,0,0.4}
\definecolor{green}{rgb}{0,0.5,0.0}
\definecolor{navy}{rgb}{0.0,0.0,0.6}
\definecolor{orange}{rgb}{0.8,0.2,0.0}
\definecolor{blue}{rgb}{0.00,0.00,1.00}

\newcommand{\msout}[1]{\text{\sout{\ensuremath{#1}}}} 
  {\color{red}}%

\usepackage[normalem]{ulem}  

\submitjournal{ApJL}

\begin{document}
\title{Oscillations of hypermassive compact stars with
  gravitational radiation and viscosity}
\correspondingauthor{Peter B. Rau}
\email{pbr44@cornell.edu}
\author{Peter B. Rau}
\affiliation{Cornell Center for Astrophysics and Planetary Science, 
Cornell University, Ithaca, NY 14850, USA
}
\author{Armen Sedrakian}
\affiliation{Frankfurt Institute for Advanced Studies,
D-60438 Frankfurt am Main, Germany}
\affiliation{Institute of Theoretical Physics, University of Wroc\l{}aw,
50-204 Wroc\l{}aw, Poland}
\begin{abstract}
  Binary neutron star mergers, such as the multimessenger GW170817
  event, may produce hypermassive compact objects which are supported
  against collapse by the internal circulation of the fluid within the
  star. We compute their unstable modes of oscillations driven by
  gravitational wave radiation and shear viscosity, modeling them as
  triaxial Riemann ellipsoids. We work in a perturbative regime,
    where the gravitational radiation reaction force is taken into
    account at 2.5--post-Newtonian order and find unstable modes with
    dissipation timescales $\gtrsim 1$ ms which are relevant to the
    transient state of a hypermassive remnant of a merger. We show that the secular
  instabilities are dominated by gravitational wave
  radiation. If the shear viscosity is
      included, it can increase  the growth times or even
    stabilize the unstable modes, but it must have values several
    orders of magnitude larger than predicted for cold neutron stars.
\end{abstract}

\section{Introduction}
\label{sec:intro}

The observation of gravitational waves (GW) from the binary neutron
star (BNS) inspiral event GW170817 by LIGO-VIRGO collaboration (LVC)
~\citep{LIGO_Virgo2017b,LIGO_Virgo2017c,LIGO_Virgo2017a} strongly
motivates the studies of the post-mergers objects left behind in such
events. Numerical simulations (for recent examples
see~\cite{Shibata2017,Most2019,Ruiz2020PhRvD,Dietrich2020}) show that
after an initial highly non-linear phase of evaluation on time-scales
of the order of 10 ms, the star settles into a gravitational
equilibrium, which is a hypermassive compact star supported against
gravitational collapse by the internal circulation of the fluid (for
reviews
see~\cite{Faber2012:lrr,Baiotti2019PrPNP,Chatziioannou2020}). The
lifetime of such an object is not well known, as it depends on several
unknown factors, such as the strength of magnetic fields or the
equation of state; however, it is expected that it could last up 100
ms and, for low-mass systems, beyond. The gravitational waves emitted
during this ``long-term'' phase of the evolution of the post-merger
object, which can be detected by the advanced LIGO, have the potential
of providing information on the integral parameters (mass, radius,
etc.) of these objects.

In this work, we report computations of the oscillation modes of
hypermassive neutron stars modeled as classical homogeneous ellipsoids
with internal circulation, i.e., {\it Riemann
  ellipsoids}~\citep{Chandrasekhar1969efe}.  In doing so, we include
the effect of shear viscosity and gravitational wave radiation, which
allows us to access the secular instabilities of these objects through
mode analysis.

Secular instabilities can develop in rapidly rotating compact objects,
the classical case being the $m=l=2$ toroidal (or bar) mode, which
becomes unstable for uniformly rotating axisymmetric stars for the
ratio of rotational kinetic energy to gravitational potential energy
$T/W \ge
0.27$~\citep{Chandrasekhar1969efe,Chandrasekhar1970ApJ}. Subsequently,
  it was shown that compact stars undergo the so-called
  Chandrasekhar-Friedmann-Schutz (CFS) instability due to
  gravitational radiation under much more general
  conditions~\citep{Friedman1978ApJ_a,Friedman1978ApJ_b}. The
  corresponding timescales for the CFS instability were studied
  (including the role played by viscosity) for Maclaurin
  spheroids~\citep{Comins1979MNRASa,Comins1979MNRASb} and rigidly
  rotating axisymmetric Newtonian models~\citep{Ipser1991ApJ}.

The case of triaxial objects with internal circulation is more
complex: the set of oscillation modes of such objects were computed in
the ellipsoidal approximation
by~\cite{Chandrasekhar1965ApJ,Chandrasekhar1966ApJ} in the
non-dissipative limit. The focus of the subsequent studies of these
objects shifted towards the problem of their secular evolution under
the action of gravitational radiation and (shear) viscosity.
\cite{Miller1974ApJ} integrated equations of motion Riemann $S$-type
ellipsoids under gravitational radiation-reaction in post-Newtonian
formalism and showed that the evolution proceeds towards either
axisymmetric or non-axisymmetric bodies without internal circulation.
Earlier, \cite{Press1973ApJ} integrated the equations of motion in the
case of viscous fluid showing that a secularly unstable, viscous
Maclaurin spheroid deforms itself into a stable, Jacobi ellipsoid,
whereby the intermediate configurations are Riemann $S$-type
ellipsoids.  \cite{Detweiler1977ApJ} integrated the relevant equations
of motion in the presence of both gravitational radiation-reaction and
viscosity. They find that the evolution ends on the axisymmetric
zero-circulation Maclaurin sequence. \cite{Lai1995ApJ} considered
gravitational wave radiation by nascent neutron stars within the
compressible ellipsoidal approximation~\citep{Lai1993ApJS} in the
presence of gravitational radiation reaction and viscosity. Their
study was mainly focused on the signatures from gravitational wave
generated by a secularly unstable new-born neutron star, although they
give analytical results for the oscillations modes of Maclaurin
spheroids. It has been acknowledged frequently in the literature
quoted above that Riemann ellipsoids are secularly unstable, however,
it appears that their oscillation modes have not been studied beyond
the non-dissipative limit given in
~\cite{Chandrasekhar1965ApJ,Chandrasekhar1966ApJ}. It is, in
  part, a purpose of this work to fill in this gap.

Below, motivated by the perspectives of observation of ``long-term''
oscillations of hypermassive neutron stars in gravitational waves we
study the spectrum of oscillation modes of Riemann
$S$-ellipsoids including the secular effects of gravitational
radiation reaction and shear viscosity in a perturbative
  manner. As the hypermassive remnants of mergers are hot, we neglect
the effects of superfluidity of their interiors, i.e. the dissipation
due to the mutual friction between the superfluid and normal fluid;
this can be treated within the ellipsoidal
approximation~\citep{Sedrakian2001PhRvD}. In this report, we
  focus only on the perturbative treatment of
  the secular effects whose growth times can be reliably computed
  using a Newtonian background. The results of the full (non-perturbative) treatment of the modes will be
  given elsewhere.

\section{Perturbation equations}

We approximate a post-merger hypermassive compact star 
 as a Riemann $S$-type ellipsoid
\citep{Chandrasekhar1969efe}, with principal axes
$a_1\neq a_2\neq a_3$ and internal circulation $\vecomega$
in the (co)rotating frame, which has angular velocity $\vecOmega(t)$
of principle axis with respect to an observer at rest. For $S$-type
ellipsoids $\vecomega$ and $\vecOmega$ are parallel, and are chosen to
lie along the Cartesian $x_3$ axis, which is the same in the inertial
and rotating frames; without loss of generality we assume
$a_1\geq a_2$.  Following \citet{Chandrasekhar1969efe} we assume
uniform density $\rho$ and incompressible fluid flow, 
$\vecnabla\cdot\vecxi=0$, where $\vecxi$ is the Lagrangian
displacement. For small perturbations, keeping terms only linear order in
displacements, and assuming time dependence of perturbations 
given by $\vecxi (t) = e^{\lambda t}\vecxi(\vecx)$
the characteristic equation is written as \citep{Chandrasekhar1969efe}
\bea
&&\lambda^2V_{i;j}-2\lambda Q_{jl}V_{i;l}-2\lambda\Omega\epsilon_{i\ell 3}V_{\ell;j} \nonumber
\\
&&-2\Omega\epsilon_{i\ell 3}(Q_{\ell k}V_{j;k}-Q_{jk}V_{\ell;k})+Q^2_{j\ell}V_{i;\ell}+Q^2_{i\ell}V_{j;\ell} \nonumber
\\
&&=\Omega^2(V_{ij}-\delta_{i3}V_{3j})+\delta\mathfrak{W}_{ij}+\delta_{ij}\delta\Pi-\delta
\mathfrak{P}_{ij}-\delta\mathcal{G}_{ij},
\label{eq:CharacteristicEquation}
\eea
where 
\begin{equation}
V_{i;j}=\int_{\mathcal{V}} d^3x\rho\xi_ix_j \label{eq:VTensorDefinition}
\end{equation}
is the perturbations of the quadrupole moment tensor,
$V_{ij} = V_{i;j}+V_{j;i}$, $Q_{ij}$ is the matrix relating the
background flow velocity inside the star $u_i$ to the coordinates in
the rotating frame $u_i=Q_{ij}x_j$.  For the Riemann $S$-type ellipsoids
with aligned spin and circulation vectors $u_1=Q_{12}x_2$,
$u_2=Q_{21}x_1$ and $u_3=0$, where
\bea
\label{eq:Q_f}
Q_{12}=-\frac{a_1^2}{a_1^2+a_2^2}\Omega f,
\quad
Q_{21}=\frac{a_2^2}{a_1^2+a_2^2}\Omega f,\quad f\equiv\omega/\Omega,
\nonumber\\
\eea
and all other elements of $Q_{ij}$ equal to zero. The expression for 
the perturbed gravitational potential energy tensor
$\delta\mathfrak{W}_{ij}$ and  pressure perturbations $\delta\Pi$ are
standard, and following \cite{Chandrasekhar1969efe} we write the
perturbation of dissipative part of the stress tensor in the ``low
Reynolds number approximation'' 
\bea
\delta\mathfrak{P}_{ij}=5\lambda\nu\left(\frac{V_{i;j}}{a_j^2}+\frac{V_{j;i}}{a^2_i}\right),
\eea
where $\nu$ is the kinematic shear viscosity, related to the dynamic shear
viscosity $\eta$ by $\nu = \eta/\rho$. Finally, the last term in
Eq.~\eqref{eq:CharacteristicEquation} is associated with the
perturbations of gravitational radiation back-reaction tensor and is
given by \citep{Miller1974ApJ}
\begin{equation}
\delta\mathcal{G}_{ij}=\frac{2G}{5c^5}\left(\msout{I}^{(5)}_{i\ell}V_{\ell j}+\msout{V}^{(5)}_{i\ell}I_{\ell j}\right),
\end{equation}
where $\msout{I}^{(5)}_{ij}$ is the fifth time derivative of the
reduced quadrupole moment tensor in the inertial
frame projected onto the rotating frame, 
\begin{equation}
\label{eq:I5}
\msout{I}^{(5)}_{ij}=I^{(5)}_{ij}-\frac{1}{3}\delta_{ij}\text{Tr}(I^{(5)}),
\end{equation}
where 
\begin{equation}
I^{(5)}_{ij}=\sum^{5}_{m=0}\sum^{m}_{p=0}C^{5}_{m}C^{m}_{p}(-1)^{p}
[(\overline{\vecOmega}^*)^p]_{ik}\frac{d^{5-m}I^{(r)}_{k\ell}}{dt^{5-m}}
[(\overline{\vecOmega}^*)^{m-p}]_{\ell j}
\label{eq:MomentofInertiaInertialFrame}
\end{equation}
and $I^{(r)}_{k\ell}$ is the moment of inertia tensor in the rotating
frame, $C^{m}_{n}$ are binomial coefficients, and for rotation about
the $x_3$ axis, the $3\times 3$ matrix
$\overline{\vecOmega}^*\equiv\Omega\vecsigma$ where the matrix
$\vecsigma$ is defined as $\sigma_{ij} = i\sigma_y$ for $ij\in 1,2$
(with $\sigma_y$ being the $y$-component of Pauli matrix) and
$\sigma_{ij} =0$ for $i=3$ or $j=3$.
The quantity $\msout{V}^{(5)}_{ij}$ is defined analogously to
\eqref{eq:I5}
\begin{equation}
\label{eq:V5}
\msout{V}^{(5)}_{ij}=V^{(5)}_{ij}-\frac{1}{3}\delta_{ij}\text{Tr}(V^{(5)}).
\end{equation}
To simplify the expressions \eqref{eq:I5} and \eqref{eq:V5} we
note that for a time-independent moment of inertia as measured in the
rotating frame, Eq.~(\ref{eq:MomentofInertiaInertialFrame}) reduces to
\begin{equation}
I^{(5)}_{ij}=\sum^{5}_{p=0}C^{5}_{p}(-1)^{p}[(\overline{\vecOmega}^*)^p]_{ik}I^{(r)}_{k\ell}[(\overline{\vecOmega}^*)^{5-p}]_{\ell j},
\label{eq:MomentofInertiaInertialFrame2}
\end{equation}
and for a triaxial ellipsoid with 
the principal axes are aligned with the coordinate axes in the
rotating frame $I_{ij}=I_{ij}^{(r)}$ 
\begin{equation}
I_{ij}=\frac{1}{5}M\delta_{ij}\left(\sum^3_{m=1}a_m^2-a_i^2\right).
\end{equation}
Now, since $\Omega$ and $I_{ij}$ are constant in time, the only
nonzero component of $\msout{I}^{(5)}_{ij}$ is \citep{Lai1994a}
\begin{equation}
\msout{I}^{(5)}_{12}=\msout{I}^{(5)}_{21}=16\Omega^2(I_{11}-I_{22}).
\end{equation}
In analogous manner we find that 
\bea
V^{(5)}_{ij}&=&\sum^{5}_{m=0}\sum^{m}_{p=0}C^{5}_{m}C^{m}_{p}(-1)^{p}[(\overline{\vecOmega}^*)^p]_{ik}\frac{d^{5-m}V_{k\ell}}{dt^{5-m}}[(\overline{\vecOmega}^*)
^{m-p}]_{\ell j} \nonumber\\
&=&\lambda^5V_{ij}-\phi_1(\lambda,\Omega)(V_{ik}(\vecsigma^4)_{kj}+\sigma_{ik}V_{k\ell}\sigma_{\ell
  j}) \nonumber\\
&+&\phi_2(\lambda,\Omega)(V_{ik}\sigma_{kj}-\sigma_{ik}V_{kj}).
\label{eq:MomentofInertiaInertialFramePerturbation}
\eea
where we have defined the auxiliary functions
\bea
\phi_1(\lambda,\Omega)&\equiv& 20\lambda\Omega^2(\lambda^2-2\Omega^2),
\\
\phi_2(\lambda,\Omega)&\equiv& \Omega(5\lambda^4-40\lambda^2\Omega^2+16\Omega^4).
\eea

\section{Characteristic equations and solution method}

The sequences of Riemann ellipsoids can be characterized by the values
of the $f$ (see Eq.~\eqref{eq:Q_f}) and the value of one of the
parameters $\alpha=a_2/a_1$ or $\beta = a_3/a_1$ (without loss of
generality $a_1 = 1$). Interesting special cases of $f$-parameter are $f=0$
corresponding to zero-circulation triaxial bodies (Jacobi ellipsoids),
$f=\pm\infty$ -- triaxial bodies supported by internal circulation
only (Dedekind ellipsoids), and $f=-2$ corresponding to an
irrotational ellipsoid in the inertial frame, since the vorticity in
this frame is given by $\vecomega_0= (2+f) \vecOmega$.

To find the modes described by Eq.~\eqref{eq:CharacteristicEquation}
we separate the nine distinct equations in the independent subsets
which are even and odd with respect to index 3.  We will not write
down the lengthly expression for each component here. We note only
that the even modes should be supplemented with
\begin{equation}
{V_{11}}{a_1^{-2}}+{V_{22}}{a_2^{-2}}+{V_{33}}{a_3^{-2}}=0,
\label{eq:IncompressibilityCondition}
\end{equation}
valid for incompressible flows. We next introduce dimensionless
quantities 
$\tilde{\Omega}\equiv{\Omega}/{\sqrt{\pi G\rho}}$,
$\tilde{\lambda}\equiv{\lambda}/{\sqrt{\pi G\rho}}$, 
and $\tilde \nu = \nu/a_1^2\sqrt{\pi G\rho}$ where $G$ is
gravitational constant. An estimate for the dimensionless kinematic
shear viscosity is
\bea\label{eq:ReducedShearViscosity} \tilde \nu =
1.35\times10^{-13}\left(\frac{\nu}{10^3\text{ cm}^2\text{
      s}^{-1}}\right)\left(\frac{\rho_0}{\rho}\right)^{1/2}\left(\frac{10\text{
      km}}{a_1}\right)^2, \nonumber\\
\eea 
where $\rho_0=2.7\times10^{14}$
g/cm$^3$ is nuclear saturation density. The timescale associated with
the damping via gravitational wave radiation 
 scales as $\tilde{t}^{5}_c$, where $\tilde{t}_c$ is
the dimensionless light crossing time
\begin{equation}
\tilde {t}_c\equiv\frac{a_1\sqrt{\pi G\rho}}{c}=0.251\left(\frac{\rho}{\rho_0}\right)^{1/2}\left(\frac{a_1}{10\text{ km}}\right).
\label{eq:ReducedLightCrossingTime}
\end{equation}
Because $\tilde \nu$ and $\tilde {t}_c$ depend on $\rho$, $\tilde{\lambda}=\tilde{\lambda}(\rho)$, unlike the case without viscosity or gravitational radiation damping in which $\lambda\propto\sqrt{\rho}$ and this $\rho$--dependence is eliminated by dividing by the characteristic frequency $\sqrt{\pi G\rho}
=7.52 \times 10^3 \times \left(\rho/\rho_0\right)^{1/2}$~Hz.

\begin{table*}
  \label{tab:1}
  \centering
  \caption{The equilibrium structure of Riemann ellipsoids for several
    values of circulation parameter $f$; note that $f=-2$ corresponds
    to irrotational and $f=0$ rigidly rotating cases. Listed are the reduced
    values of semi-axis $\alpha=a_2/a_1$ and $\beta=a_3/a_1$ and the non-dimensional
    rotation frequency $\tilde{\Omega}$.  }
\begin{tabular}{c|cc|cc|cc|cc|cc}
  \hline
$\alpha$ &  $\beta$ & $\tilde \Omega$ &  $\beta$ & $\tilde\Omega$ &  $\beta$ & $\tilde\Omega$ &
                                                                    $\beta$
  & $\tilde\Omega$ &  $\beta$ & $\tilde\Omega$\\
  \hline
  &  $f = -3$ &  &  $f = -2$ &   &   $f = 0$ &  &  $f
                                                              = 2$ & &
                                                                       $f = 3$
  \\

  \hline
1.0 & 0.96 & 0.160 &   1.00 & 0.267 &   0.58 & 0.374 &   0.30 & 0.110 &   0.32 & 0.071 \\
0.9 & 0.91 & 0.161 &   0.95 & 0.267 &   0.55 & 0.373 &   0.29 & 0.110 &   0.30 & 0.071 \\
0.8 & 0.86 & 0.163 &   0.88 & 0.268 &   0.52 & 0.367 &   0.27 & 0.111 &   0.28 & 0.072 \\
0.7 & 0.79 & 0.167 &   0.81 & 0.269 &   0.48 & 0.356 &   0.25 & 0.112 &   0.26 & 0.073 \\
0.6 & 0.72 & 0.172 &   0.73 & 0.269 &   0.43 & 0.337 &   0.23 & 0.114 &   0.24 & 0.075 \\
0.5 & 0.63 & 0.179 &   0.63 & 0.266 &   0.38 & 0.310 &   0.21 & 0.116 &   0.21 & 0.077 \\
0.4 & 0.53 & 0.184 &   0.52 & 0.253 &   0.33 & 0.270 &   0.19 & 0.117 &   0.18 & 0.080 \\
0.3 & 0.41 & 0.178 &   0.38 & 0.221 &   0.26 & 0.215 &   0.17 & 0.114 &   0.15 & 0.082 \\
0.2 & 0.26 & 0.142 &   0.24 & 0.156 &   0.18 & 0.144 &   0.14 & 0.097 &   0.12 & 0.077 \\
0.1 & 0.12 & 0.064 &   0.11 & 0.065 &   0.10 & 0.060 &   0.09 & 0.052 &   0.08 & 0.047 \\
\hline
\hline
\end{tabular}
\end{table*}

We first obtain the equilibrium sequences of Riemann ellipsoids
characterized by $\alpha$, $\beta$, $\tilde{\Omega}$ for fixed
$f$. The equilibrium semi-axis values ($a_1 = 1)$ and rotation
frequency $\tilde{\Omega}$ for representative cases are listed in
Table~\ref{tab:1}.   Then the characteristic frequencies are
computed separately for the even and odd modes. For the even modes, we
eliminate the pressure perturbations and obtain a system of five
equations with six unknowns $V_{11}$, $V_{22}$, $V_{33},$ $V_{1;2}$,
$V_{2;1}$ and $\tilde{\lambda}$, which can be solved for
$\tilde {\lambda}$ by first writing it as a matrix equation
\begin{equation}
\mathbf{M}_{\text{even}}(\tilde\lambda)\cdot(V_{11}, V_{22}, V_{33}, V_{1;2}, V_{2;1})^{\text{T}}=0,
\label{eq:EvenModesMatEq}
\end{equation}
where $\text{T}$ indicates the tranpose, and demanding that the determinant of the matrix
$\mathbf{M}_{\text{even}}$ is zero. The resulting characteristic equation is a polynomial in
$\tilde{\lambda}$ of order 17 in the dissipative case, whereas it is
of order 8 in the absence of dissipation (i.e. both viscosity and
gravitational radiation). Note that for the Riemann $S$-type ellipsoids
four of the roots are zero in the non-dissipative case, whereas only
one root is zero when shear viscosity and gravitational radiation
damping are included. Similarly, for odd modes we find 
\begin{equation}
\mathbf{M}_{\text{odd}}(\tilde{\lambda})\cdot(V_{1;3},V_{3;1},V_{2;3},V_{3;2})^{\text{T}}=0,
\label{eq:OddModesMatEq}
\end{equation}
which leads to a characteristic equation which is a polynomial in
$\tilde{\lambda}$ of order 14; in the non-dissipative case 
this reduces to a polynomial of order 8. In discussing the numerical
results for the modes, we define $\tilde{\sigma}=-i \tilde{\lambda}$ and 
an associated (dimensionless) dissipation timescale 
\begin{equation}
  \tilde{\tau}=-\frac{1}{\text{Im}(\tilde{\sigma})} 
\end{equation}
Thus, for $\text{Im}(\sigma)<0$ the modes are unstable with
characteristic (dimensional) growth time $\tau = (
\pi\rho G)^{-1/2}\tilde \tau$. 

\section{Numerical results}

To obtain physical values of oscillation frequencies and their damping
or growth time-scales we need to evaluate the quantities
$\tilde {\nu}$ and $\tilde {t}_c$, i.e., we need to specify
physically motivated values of $a_1$, $\rho$ and $\nu$, as well as the mass and the radius of the object. To model a hypermassive
compact object resulting from a merger of two neutron stars we choose
the value $2.74M_{\odot}$ corresponding to the GW170817
event~\citep{LIGO_Virgo2017b,LIGO_Virgo2017c,LIGO_Virgo2017a}. We set
the uniform density $\rho=3.62\rho_0$ by enforcing that the $f=-2$,
$\alpha=\beta=1$ star has a radius of $a_1=11$ km. This is a reasonable value for the radius of the semi-homogeneous core of the star-- we assume that an outer envelope region of thickness 1--2 km is unimportant for the oscillations. We adjust $a_1$ so that each
ellipsoid has constant volume and hence the same mass. This results
in both $\tilde{\nu}$ and $\tilde{t}_c$ varying accordingly.

If we assume that the viscosity of the hypermassive neutron star is
due to the (non-superfluid) core (consisting of neutrons, protons and
charged leptons in beta equilibrium) then viscosity is dominated by the
normal neutron fluid. Taking as a reference value the low-temperature
results of ~\cite{Flowers1981ApJ,Shternin2013}
$\eta \sim 10^{19}/T_8^{2}$ g~cm$^{-1}$~s$^{-1}$ for temperature
$T = 10^8$ K and $\rho = \rho_0$, we find that the kinematic viscosity
would be of the order of $\nu \sim 4$ cm$^{2}$~s$^{-1}$ for
characteristic temperature of a hypermassive neutron star 1~MeV. A kinematic viscosity of this magnitude is far too
small compared to the dissipation via gravitational-wave radiation.
If the temperature of a hypermassive neutron star is above the
neutrino trapping temperature $T\le  10$~MeV, then neutrinos can
contribute to the viscosity.  This contribution can be estimated using 
the well-known kinematical formula for classical gases
\begin{equation}
\label{eq:eta}
    \eta = \frac{1}{5}n_{\nu}\ell_{\nu}p_{\nu}
\end{equation}
where $\ell_{\nu}$, $n_{\nu}$, $p_{\nu}$ are the neutrino mean free
path, number density, and momentum respectively. If neutrinos are
close to the ballistic regime (in the vicinity of trapping
temperature), then we can take $\ell_{\nu}\simeq 10$ km -- the size of
the system. Approximating further $n_{\nu}\simeq 0.1\rho/m_N$ where
$m_N$ is the nucleon mass, $p_{\nu} \simeq 10$ MeV$/c$, we find
$\nu\simeq 10^{13}$ cm$^2$s$^{-1}$ and hence
$\tilde{\nu}\sim10^{-3}$. At this size, the viscosity starts to have a
noticeable effect on the instability growth times [see
Fig.~(\ref{fig:2})]. We use an artificially enhanced kinematic
viscosity $\nu^* = 10^{14}$ cm$^2$s$^{-1}$; given the simplistic
nature of \eqref{eq:eta}, which does not take into a number of
physical factors (e.g. turbulent viscosity), such a choice is not
prohibitive. It has been argued that bulk viscosity can be the
  dominant mode of dissipation in hot compact
  stars~\citep{Sawyer1989PhRvD} and neutron star
  mergers~\citep{Alford2019PhRvC,Alford2019PhRvD}, which however would
  require a treatment beyond the incompressible limit adopted here.
\begin{figure}[tb]
    \centering
    \includegraphics[width=\columnwidth]{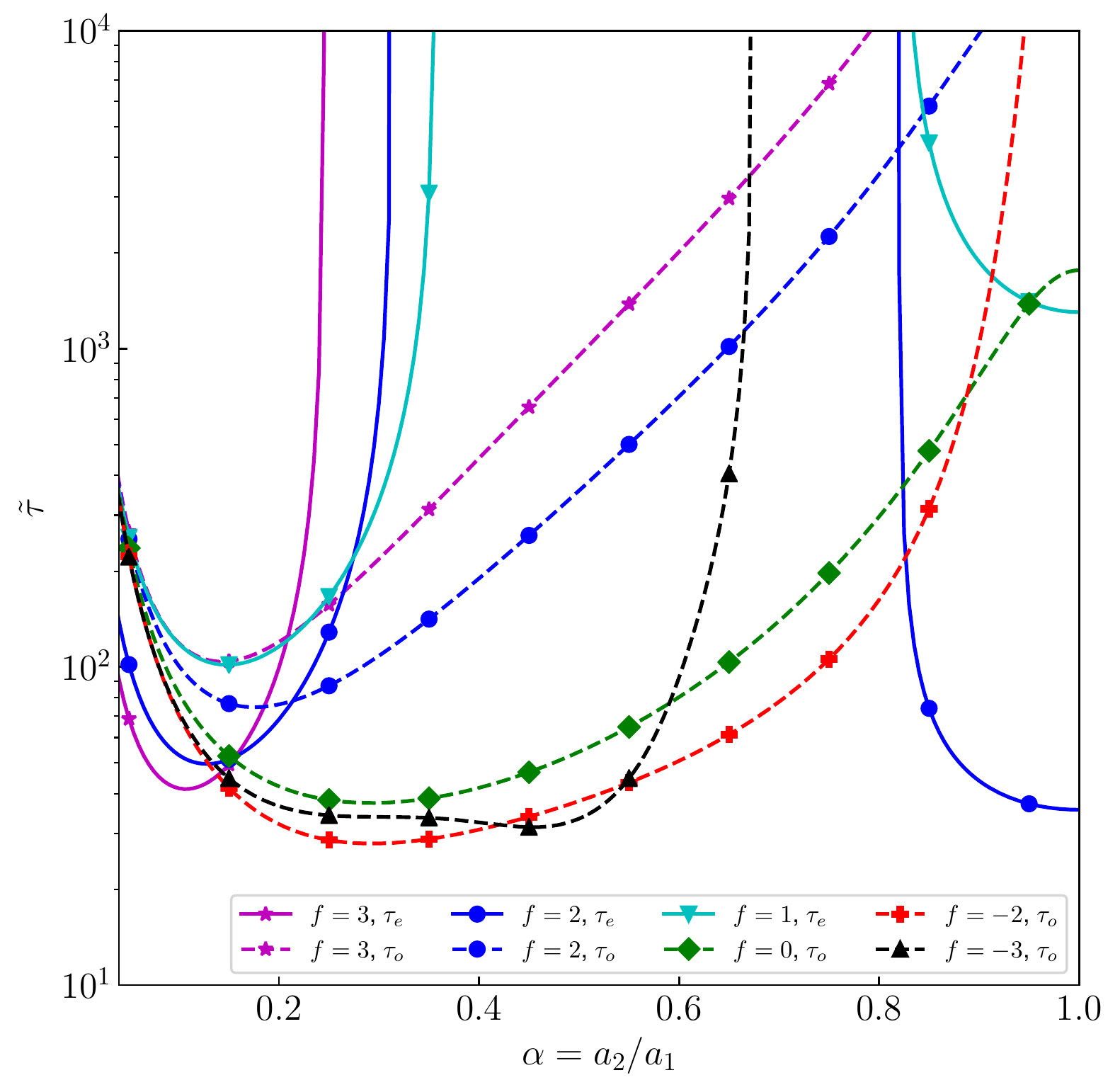}
    \caption{ The dimensionless growth times for the perturbative unstable modes with zero shear viscosity
      for $f=3,2,0,-2,-3$ (odd modes) and $f=3,2,1$ (even modes) for
      stellar model $M=2.74M_{\odot}$ as a function of
      $\alpha=a_2/a_1$. The even/odd modes are distinguished by
      solid/dashed lines respectively. The dimensional growth
      timescale in units of ms is given by
      $\tau = (\tilde \tau/7.52) \left(\rho/\rho_0\right)^{-1/2}$.  }
    \label{fig:1}
\end{figure}
\begin{figure}[tb]
    \centering
    \includegraphics[width=\columnwidth]{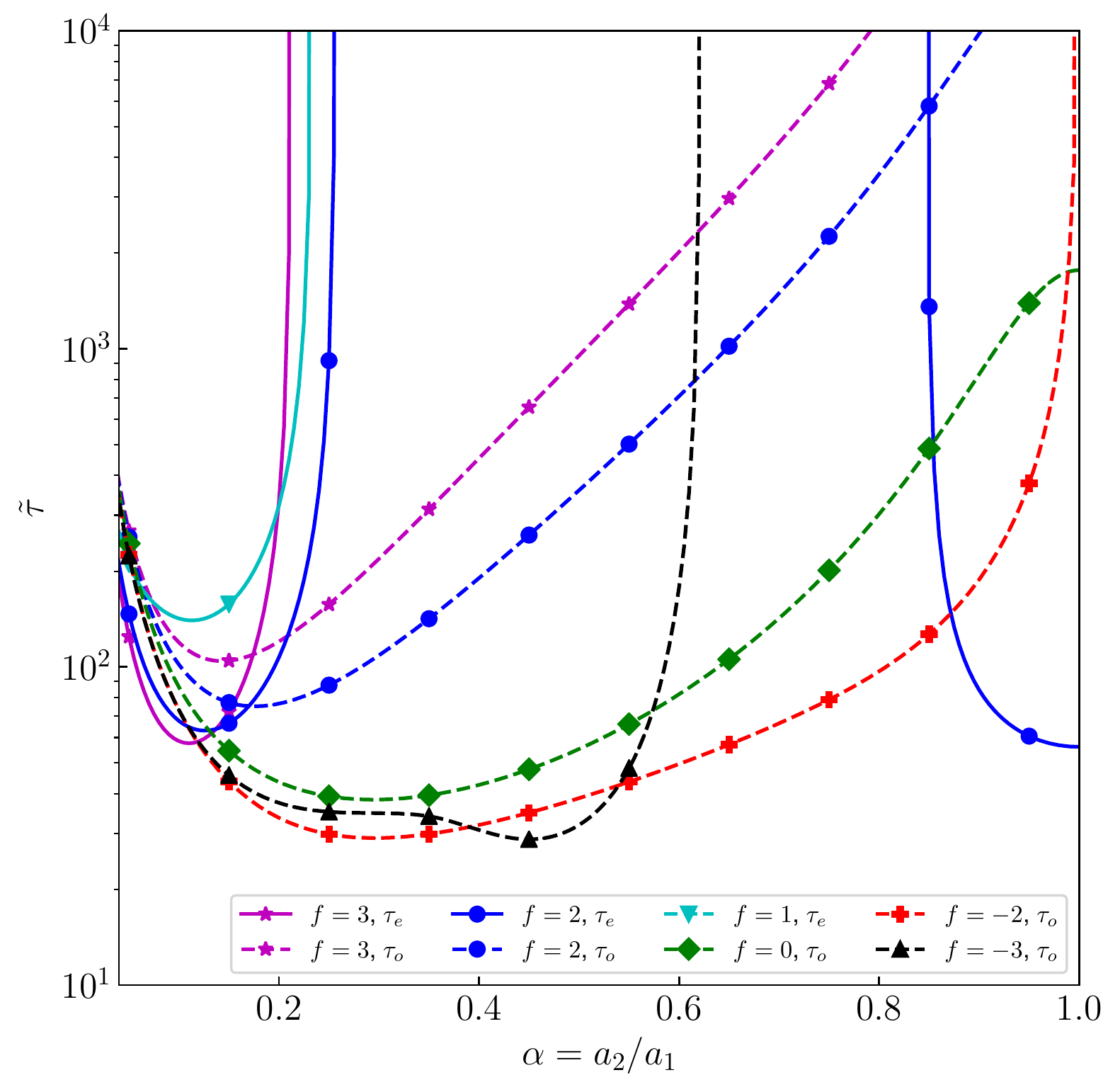}
    \caption{Same as in Fig.~\ref{fig:1}, but including
      (artificially enhanced) viscosity $\nu^*=10^{14}$ cm$^2$s$^{-1}$. }
    \label{fig:2}
\end{figure}

It has been shown by ~\citet{Chandrasekhar1969efe} (see $\S$49) that
Riemann $S$-type ellipsoids undergo dynamical (i.e. in the absence of
dissipation) instability by one of the odd-parity modes when $f<-2$.
Otherwise, the remaining odd and even parity modes were found to be
stable.  By explicitly solving the characteristic equations for the
odd and even parity modes, we find that the Riemann $S$-type
ellipsoids are generically unstable when gravitational radiation and
shear viscosity are included. In doing so we extract those
  modes that are perturbative, in the sense that they have an oscillatory
part approximately equal to that of an undamped mode, i.e., their
complex frequencies can be expressed as $\sigma=\sigma_0+\Delta\sigma$
where $\sigma_0\in\mathbb{R}$ is an undamped mode frequency and
$\Delta\sigma\in\mathbb{C}$ is a small correction.  This is in agreement with the
computation of gravitational radiation-induced unstable toroidal modes
of Maclaurin spheroids by~\citet{Chandrasekhar1970ApJ}. The
  perturbative treatment is required for two reasons: (a) the
  gravitational radiation back-reaction force we use is computed to
  2.5--post-Newtonian order whereas our background ellipsoids are
  Newtonian, i.e., post-Newtonian corrections to the background should
  be added for consistency on the hydrodynamical timescales; (b) the
  secular effects on the hydrodynamical scales might require a more
  complete form of the gravitational radiation back-reaction force
  given by \citet{Chandrasekhar1970a}, than the one employed here
  \citep{Miller1974ApJ}; these two forms agree in the perturbative
  limit.

Turning to the results of the mode analysis, we show in
Fig.~\ref{fig:1} the growth time-scales of unstable odd and even modes
for various $f$ values in the absence of viscosity. The growth times are labelled $\tau_{e}$ for even modes and $\tau_{o}$ for odd modes. It is seen
that there is one unstable odd mode for each $f$, and that there are even unstable modes
only for $f > 0$, with one such mode for each $f$. In the special case $f=0$ and $\alpha=1$, when the 
Riemann S-type ellipsoid reduces to a Maclaurin spheroid, there is a
point of marginal stability for eccentricity $e=0.81267$ where
$\text{Im}(\sigma)=0$ for one of the even modes, in agreement with
\citet{Chandrasekhar1970}. In general, the unstable odd modes can
have shorter growth times than the unstable even modes and are
unstable for a greater subset of the stable Riemann S-type
ellipsoids. The unstable even $f=1,2$ modes start off unstable for
large $\alpha$, then are first stabilized and subsequently become
unstable again as $\alpha$ decreases. 

The effect of (artificially enhanced) viscosity
  $\nu^*=10^{14}$~cm$^2$~s$^{-1}$ on the growth times of unstable
  modes are shown in Fig.~\ref{fig:2} for several values of
  $f$. It is seen
    that viscosity has little effect on the growth times of the
  unstable odd modes. For the even modes, viscosity
  suppresses the gravitational radiation-induced instability and
  increases the corresponding growth times, in some cases moving them
  outside of the physical relevant range. This suppression is in
  agreement with previous results found for axisymmetric Newtonian
  stars~\citep{Ipser1991ApJ}. 
\begin{figure}[tb]
    \centering
    \includegraphics[width=\columnwidth]{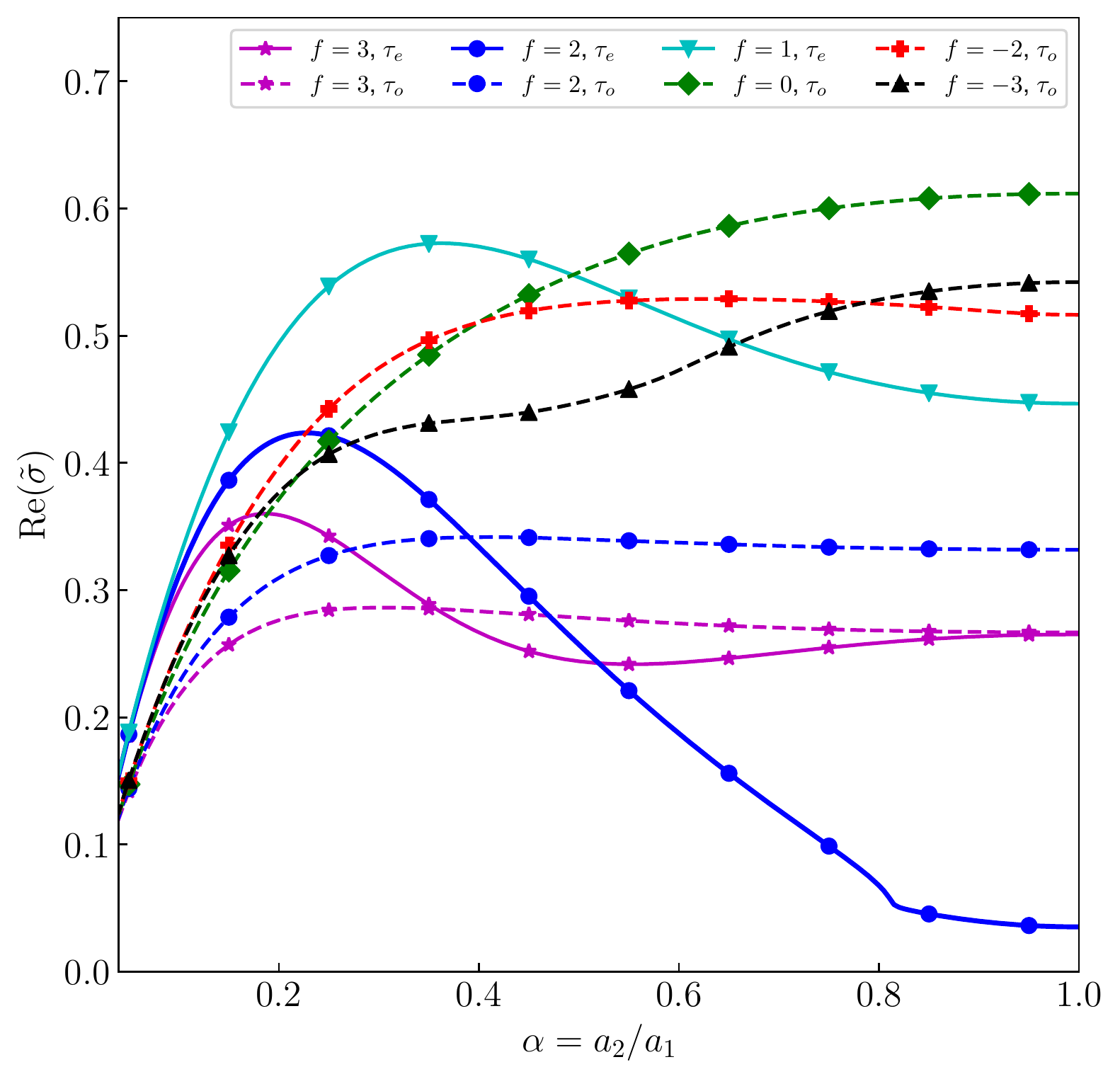}
    \caption{ The oscillation frequencies $\rm{Re}(\tilde{\sigma})$
      for the ``slow'' unstable modes with zero shear viscosity for
      $f=3,2,0,-2,-3$ (odd modes) and $f=3,2,1$ (even modes) for
      stellar model $M=2.74M_{\odot}$ as a function of
      $\alpha=a_2/a_1$. The line markers used for each mode match
      those used for the growth times $\tilde{\tau}$ of that mode in
      Figure~\ref{fig:1}.  }
\label{fig:3}
\end{figure}
  Figure~\ref{fig:3} show the dimensionless oscillation frequencies
  $\rm{Re}(\tilde{\sigma})$ of unstable modes in the absence of
  viscous dissipation. The changes due to the viscosity are
  insignificant (Note that the growth timescale and oscillation
  frequencies in Figs.~\ref{fig:1}--\ref{fig:3} are shown with the
  same line markers). The dimensionless frequencies of the modes are
  concentrated mostly the range
  $0.1\le {\rm Re}(\tilde{\sigma}) \le 0.6$, which given the
    value of the normalization frequency
    $\Omega_0 = (\pi \rho G)^{1/2} = 7.52 \times 10^3
    \left(\rho/\rho_0\right)^{1/2}$~s$^{-1}$, correspond to
  frequencies in the kHz range.

\section{Summary}
It is anticipated that hypermassive neutron stars left behind by a
binary neutron star merger will be detectable by advanced
gravitational wave detectors at kHz frequencies. After a short
transient, the star is expected to settle down in an equilibrium
configuration supported by internal circulation (provided there is no
prompt collapse to a black hole). Here we have modeled a hypermassive
neutron star by a classical Riemann ellipsoid - rotating triaxial body
supported by internal circulation - and derived the perturbative set of small-amplitude oscillation modes taking
into account the dissipation through gravitational wave radiation and
viscosity. In general, the obtained characteristic equations
  have 17 even parity and 14 odd parity modes, among which we
  identified the class of perturbative unstable
  modes with growth timescales of the order of
  $\gtrsim1$~ms and eigenfrequencies in the kHz range (see
  Figs.~\ref{fig:1}-\ref{fig:3}). The instability of the modes is due
  to the gravitational radiation. Adding (artificially
  enhanced) viscosity suppresses the instability of the modes,
  increasing their growth times, in some cases, beyond the physically
  interesting regime. 
  
 A prerequisite of our analysis is that the star
is in an approximate equilibrium, which can be tested only through
numerical simulations of the post-merger transient. This implies that the growth times of unstable modes are shorter than the timescales over which the unperturbed equilibrium changes appreciably, which is the case, for example, for models of Riemann ellipsoids computed in~\cite{Miller1974ApJ}. Provided that the quasi-equilibrium state has been reached, our analysis accounts for the complete set of the dissipative modes of a hypermassive neutron star within approximations employed.

Our modeling of hypermassive neutron stars can and should be improved
in the future by adding more realistic features, such as non-uniform
matter distribution, realistic equations of state, and effects of
relativity in describing the fluid perturbations and the
  equilibrium stellar models. However, the current insights into the
instabilities that develop in Riemann $S$-type ellipsoids could be of
interest also in more general context of stellar equilibria and
oscillations, physics of nuclei as well as trapped atomic clouds.

\section*{Acknowledgements}

We thank Ira Wasserman for discussions and the referee for many useful comments.
P. B. R. acknowledges the support of the Boochever Fellowship for spring 2020
and the hospitality of Frankfurt Institute for
Advanced Studies. A.S.\ acknowledges the support by
the DFG (Grant No.  SE 1836/5-1) and the European COST Action
``PHAROS'' (CA16214).

\bibliographystyle{aasjournal}
\bibliography{Postmerger_ref,GW_ref,library,textbooks}	

\end{document}